\documentclass[aps,prd,preprint,superscriptaddress,showpacs,nofootinbib]{revtex4-1}

\usepackage{color}
\usepackage{framed}
\usepackage{amsmath}
\usepackage{amssymb}
\usepackage{graphicx}
\usepackage[normalem]{ulem} 
\usepackage{times}

\def\del{\partial}

\begin{document}

\title{
Composite and elementary natures of $a_1(1260)$ meson
} 

\author{Hideko Nagahiro}
\thanks{}
\affiliation{Department of Physics, Nara Women's University, 
Nara 630-8506, Japan}
\affiliation{Research Center for Nuclear Physics (RCNP), Osaka
University, Ibaraki, Osaka 567-0047, Japan}

\author{Kanabu Nawa}
\author{Sho Ozaki}
\affiliation{Research Center for Nuclear Physics (RCNP), Osaka
University, Ibaraki, Osaka 567-0047, Japan}

\author{Daisuke Jido}
\affiliation{Yukawa Institute for Theoretical Physics, Kyoto University, 
Kyoto 606-8502, Japan}

\author{Atsushi Hosaka}
\affiliation{Research Center for Nuclear Physics (RCNP), Osaka
University, Ibaraki, Osaka 567-0047, Japan}

\date{\today}

\begin{abstract}
We develop a practical method to analyze the mixing structure of hadrons
 consisting of two components of quark composite and hadronic composite.  As an
 example we investigate the properties of the axial vector meson
 $a_1(1260)$ and discuss its mixing properties quantitatively.  We also
 make reference to the large $N_c$ procedure and its limitation for the
 classification of such a mixed state. 
\end{abstract}

%\keywords{}

\maketitle

%\section{}
%\subsection{}
%\subsubsection{}
%
\section{Introduction}

Hadrons, interacting with the strong force, are composite objects 
of quarks and gluons. One of recent interests in the hadron structure 
is whether hadrons are made up of quarks and gluons confined in a
single-particle potential as described in the conventional quark 
model, or rather develop subcomponents of quark-clusters inside hadrons. A typical 
example of the latter is the deuteron, which is composed
of a proton and a neutron not by six quarks in a single confining
potential~\cite{Weinberg:1962hj,*Weinberg:1965zz}.  It has been also suggested that 
some hadronic resonances could have substantially large components 
of hadronic composites~\cite{Baru:2003qq,Hyodo:2008xr}. 
Such recursive (nesting-box) structures are also seen in nuclear physics.
For instance, the first $0^{+}$ excited state of $^{12}$C is described 
as a three-$\alpha$ cluster state~\cite{Funaki:2003af}. 

In general it is not easy to clearly identify the quark-cluster
components, because the strong interaction scales for quarks and hadrons
are not well separated.  To simplify the situation, it would be a good
starting point to set up a model space of a state described by a single-particle
potential and the one of several quark-clusters.  The former may be 
identified with the ``elementary'' component\footnote{The elementary
component was referred to as the CDD pole or genuine quark state in the
literature~\cite{Hyodo:2008xr}.} 
while the latter with the
(hadronic) composite, the situation which was studied by
Weinberg~\cite{Weinberg:1962hj,*Weinberg:1965zz}. 
If
hadronic resonant states are unavoidably
mixture of hadronic and quark-composites, 
an important issue is to clarify how these 
components are mixed in a hadron.

One good example to study the two features and their mixing is
provided by 
the low-lying axial 
vector meson $a_1(1260)$.
The $a_1$ meson is a candidate of the chiral partner of the
$\rho$ meson~\cite{Weinberg:1967kj,Bernard:1975cd,Ecker:1988te} described as a
$q\bar{q}$-composite, for example, in the Nambu-Jona-Lasinio 
model~\cite{Dhar:1983fr,Hosaka:1990sj} and 
in the Lattice calculation~\cite{Wingate:1995hy}.
It can also appear as a gauge boson of the hidden local
symmetry~\cite{Bando:1987br,*Kaiser:1990yf}, which is recently reconciled
with the five-dimensional gauge field of the holographic
QCD~\cite{Sakai:2004cn,*Sakai:2005yt,Nawa:2006gv}.
On the other hand, 
in coupled-channel approaches based on the chiral effective
theory~\cite{Roca:2005nm,Lutz:2003fm}, the $a_1$ meson has been
described as a dynamically generated resonance in the $\pi\rho$
scattering without introducing its explicit pole term{.} 
{T}he $a_1$ nature {has been studied} by calculating
physical observables such as the radiative decay
width~\cite{Roca:2006am,*Nagahiro:2008zza,*Nagahiro:2008cv} or the $\tau$
decay spectrum into three pions~\cite{Wagner:2008gz,GomezDumm:2003ku,*Dumm:2009va}.
Yet, the {internal structure} of the $a_1$
meson is not well understood. 
%{Similar examples are also given by many systems, for
%instance, the $f_0(600)$ (or 
%$\sigma$), $f_0(980)$ and $N^*(1535)$
%resonances~\cite{Hyodo:2010jp,Caprini:2005zr,Jido:2002yb,*Nagahiro:2003iv}.
%}

In this {paper}, we {focus on} hadron structure having two 
components of 
quark-composite (we refer to it as the elementary component)
 and hadronic composite. {We} propose a method {to disentangle their
mixture appearing in the 
physically observed state,
by taking the $a_1$ meson as an example.   
{The method provided here can be generally
applied to other
mixed systems if the interaction {is} given between 
{the}
elementary component and possible constituents
making {the} composite state.}
Our ingredients for the study of $a_1$ are therefore
$\pi$ and $\rho$ mesons\footnote{We regard the $\rho$ meson as
stable particles in the present model setting.} which has potential to
generate the {\it composite} $a_1$ meson with a suitable $\pi \rho$
interaction, and the {\it elementary} component of the $a_1$ meson 
(dominated by $q \bar q$) which couples to the $\pi\rho$ pair by
a three-point interaction.
{We first solve the $\pi \rho$ scattering amplitude to find the poles
corresponding to the physical $a_1$, and
then develop a method to clarify the mixing nature of the two
components by introducing appropriate {bases}
for pure composite  
and elementary components.
As an extension, we apply our method to the study of
the large $N_c$ behavior of the $a_1$
state.  The large $N_c$ limit is usually considered to give a
guiding principle for the classification of state, in which the mass of
the $q\bar{q}$ bound state scales as ${\cal O}(1)$ {in powers
of $N_c$} while 
that of a meson-meson molecule scales with higher order of 
$N_c$~\cite{Witten:1979kh,Jaffe:2007id}.  
{Reference \cite{Pelaez:2009}, however, has brought
caution for the use of large $N_c$ argument for the dynamically 
generated scalar resonances.}
We discuss the validity
of the classification for
the mixed states of hadronic composite and elementary
components.
%\footnote{Reference \cite{Pelaez:2009} has also brought
%caution for the use of large $N_c$ argument for the dynamically 
%generated scalar resonances.}

\section{Formalism\label{sec:formalism}}
%\subsection{Scattering amplitude}
Let us start with the composite $a_1$ meson which is dynamically
generated in the $s$-wave $\pi\rho$ scattering
through the non-perturbative dynamics.
The scattering
amplitude $t$ satisfies the 
{Bethe-Salpeter} equation, 
%\begin{equation}
$t=v+vGt$, 
%\end{equation}
where $v$ is a four-point $\pi\rho$ interaction and $G$ the $\pi\rho$ two-body
propagator and its formal solution is given by,
\begin{equation}
t=\frac{v}{1-vG}\ .
\label{eq:T_pirho}
\end{equation}
If the potential $v$ is sufficiently attractive, the 
amplitude develops a pole corresponding to a composite bound or resonant state of
the scattering system
{at the energy satisfying $1-vG=0$}.
In the present {case of} $a_1$, the potential $v$ can be
obtained from the $s$-wave projection of the
Weinberg-Tomozawa interaction~\cite{Roca:2005nm},
and the pole in the $\pi\rho$ scattering
amplitude can appear above the $\pi\rho$ threshold as a resonance
in the second Riemann sheet,
{which is a consequence of the energy dependent interaction.}
This {pole} corresponds to the $\pi\rho$-composite
$a_1$ meson~\cite{Roca:2005nm} without $q\bar{q}$
quark-core~\cite{Hyodo:2008xr,Lurie:1964,*Lurie:book}.

Because the elementary $a_1$ meson has a coupling to the $\pi$ and
$\rho$ mesons, it also contributes to the $\pi \rho$ scattering
amplitude {in the form of an} 
effective $\pi \rho$ interaction going through the elementary $a_1$
pole: 
\begin{equation}
v_{a_1}=g\frac{1}{s-m_{a_1}^2+i\epsilon}g
\label{eq:v_a1},
\end{equation}
where $g$ is the coupling to $\pi \rho$ that can depend on $s$, and
$m_{a_1}$ the bare mass  
of the elementary $a_1$ meson. 
The full scattering amplitude $T$ having both the $\pi\rho$ four-point 
interaction $v$ and the $a_1$ pole term $v_{a_1}$ is then written by,
\begin{equation}
T=\frac{v+v_{a_1}}{1-(v+v_{a_1})G}\ .
\label{eq:T-matrix}
\end{equation}
This amplitude generates poles corresponding to
{physical} resonant states of the problem.  They are expressed as a superposition
of the basis states associated with the two poles {of} composite $a_1$ in
Eq.~(\ref{eq:T_pirho}) and 
elementary $a_1$ in Eq.~(\ref{eq:v_a1}), respectively.

Now let us study the mixing nature of the physical states. To this end,
we first {express equivalently}
the amplitude $t$ in Eq.~(\ref{eq:T_pirho}) as, 
\begin{equation}
t \equiv g_R(s)\frac{1}{s-s_p}g_R(s),
\label{eq:T_WT}
\end{equation}
where $s_p$ is the pole position of  
the amplitude $t$ in Eq.~(\ref{eq:T_pirho}).  In this form, we can
interpret 
$(s-s_p)^{-1}$ as the one-particle propagator of the composite $a_1$
meson  as shown in
Fig.~\ref{fig:WT_pole} by taking an analogy with the conventional discussion of bound
state problem~\cite{Lurie:1964,*Lurie:book}.  The vertex function $g_R(s)$ 
that is defined so as to reproduce Eq.~(\ref{eq:T_pirho}) exactly is
interpreted as the effective coupling of the composite $a_1$ to
$\pi\rho$.  This interpretation works well in the neighborhood of the
pole, $s\sim s_p$.  As $s$ is further away from $s_p$, $g_R(s)$ receives 
more contributions from the non-resonant background~\cite{Doring:2009bi}.

\begin{figure}[hbt]
\includegraphics[width=0.8\linewidth]{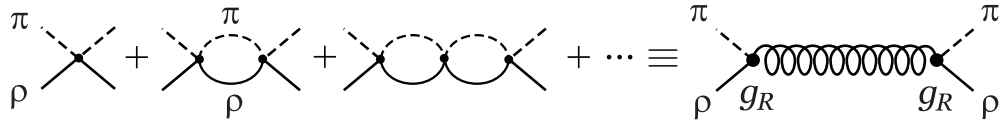}% Here is how to import EPS art
\caption{Infinite set of diagrams that contributes to the meson-meson
 scattering amplitude in Eq.~(\ref{eq:T_pirho}) and the definition of the propagator
 of the composite $a_1$ meson in Eq.~(\ref{eq:T_WT}).
\label{fig:WT_pole}
}
\end{figure}

Having the form of Eq.~(\ref{eq:T_WT}),  we {now} rewrite the
scattering amplitude $T$ in 
Eq.~(\ref{eq:T-matrix}) as,
\begin{equation}
T=(g_R,g)\frac{1}{\hat{D}_0^{-1}-\hat{\Sigma}}
\begin{pmatrix}g_R\\g\end{pmatrix}\ ,
\label{eq:T-matrix2}
\end{equation}
where
\begin{equation}
\hat{D}_0^{-1}=
\begin{pmatrix}
s-s_p
& \\
& s-m_{a_1}^2
\end{pmatrix}, \ \ 
\hat{\Sigma}=
\begin{pmatrix}
&g_RGg \\
gGg_R & gGg
\end{pmatrix}\ .
\label{eq:D0}
\end{equation}
The diagonal elements of the matrix $\hat{D}_0$ are the free
propagators of the two $a_1$'s, one for the composite and the other
for the elementary ones having the proper normalization,
and the matrix $\hat{\Sigma}$ expresses the self-energy and interactions
for these modes.
One can prove that Eq.~(\ref{eq:T-matrix2}) is {identical} with
Eq.~(\ref{eq:T-matrix})  
after some algebra. 

We emphasize that the expression of Eq.~(\ref{eq:T-matrix2}) makes it
possible to analyze the mixing nature of the physical $a_1$ in terms of
the original {two bases}.
Having the amplitude in the form of Eq.~(\ref{eq:T-matrix2}),
{the matrix}
$
\hat{D}{\equiv}({\hat{D}_0^{-1}-\hat{\Sigma}})^{-1}
$
{is identified with the propagators of the physical states
represented by the bases of the elementary and composite $a_1$'s.}
The diagonal elements $D^{ii}$ indicate the dressed
propagators of the composite and the elementary $a_1$'s as shown in
Fig.~\ref{fig:full_D}, which express the $a_1$ mesons acquiring the
quantum effects: e.g., 
\begin{figure}[h]
\includegraphics[width=0.8\linewidth]{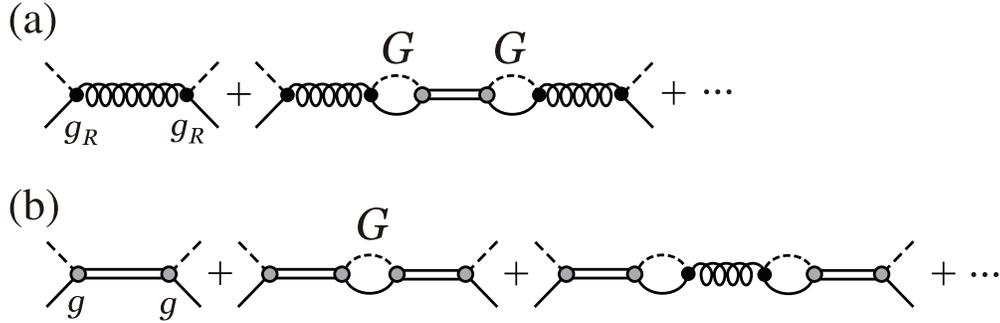}% Here is how to import EPS art
\caption{
{Diagrams that contribute to the full amplitude in
 Eq.~(\ref{eq:T-matrix2}) going through (a) the component of the dressed
 propagator 
 $D^{11}$ and 
 (b) that of $D^{22}$. The dashed and solid internal lines
 indicate the $\pi$ and $\rho$ propagators, while the curved and double
 lines are those of the composite and elementary $a_1$'s.}
\label{fig:full_D}}
\end{figure}
the elementary $a_1$ introduced without width in
Eq.~(\ref{eq:v_a1}) now has a width owing to its decay into $\pi\rho$ system.
The important features of the propagators are that they have poles exactly
at the same positions as the full
amplitude $T$ in Eq.~(\ref{eq:T-matrix}), and the 
residues of the diagonal elements
$D^{ii}$ defined by,
\begin{equation}
{D}^{ii}=\frac{z_a^{ii}}{s-M_a^2}+\frac{z_b^{ii}}{s-M_b^2}
+(\text{regular term})
\ 
(i=1,2)\ ,
\label{eq:Res}
\end{equation}
have {the meaning of the wave function renormalization} and {then}
carry the information on the mixing rate of the physical resonant states.
For instance, the residue $z_a^{11}$ means the probability of finding
the original composite $a_1$ component in the resulting state having the
mass $M_a$.  

The above discussions have close analogy to a two-level problem, which
is well realized when the energy $s$ is not 
very far away from the original two poles $s_p$ and $m_{a_1}^2$.  As $s$ is
getting away from them, the orthogonality of the physical two resonant
states is lost and furthermore, the residues can exceed unity.  These
problems arise owing to the energy dependence of $\hat{\Sigma}$,
especially those of $g_R$ and $g$ in $\hat{\Sigma}$.
We note that the energy dependence of $g_R$ is unavoidable for
the resonant states generated dynamically.

\section{Application to the $a_1(1260)$ system}
So far, we have given a general framework to investigate the mixing
properties of hadrons consisting of multiple components having different origins.
To perform a concrete calculation, we shall employ a suitable model for
$a_1$.  Here we take only the single $\pi\rho$ channel into account for
the composite $a_1$ because other channels have been found not
important~\cite{Roca:2005nm}, but we can easily extend our framework 
in Sec.~\ref{sec:formalism}
to
coupled-channel cases.  
Our ingredients for the study of the $a_1$ system are, therefore, the
$\pi$, $\rho$ and (elementary) $a_1$ mesons,
and for which suitable models are given.

In this paper, as for interaction Lagrangians, we adopt the chiral Lagrangians
induced by the 
holographic QCD {approach} as the Sakai-Sugimoto
model~\cite{Sakai:2004cn,*Sakai:2005yt}, 
% with D4/D8/$\overline{{\rm D}8}$ multi-D brane system in
%type IIA superstring theory~\cite{Sakai:2004cn,*Sakai:2005yt}, 
that is 
constructed in the strong-coupling limit of large-$N_c$ QCD.
The relevant
Lagrangians~\cite{Sakai:2004cn,*Sakai:2005yt} are given by,
\begin{gather}
{\cal L}_{\rm WT}=\frac{1}{f_\pi^2} {\rm tr}
([\rho^\mu,\del^\nu\rho_\mu][\pi,\del_\nu\pi]
)\ ,
\label{eq:L_WT}
\\
\begin{split}
{\cal L}_{a_1\pi\rho}=-ig_{a_1\pi\rho}\frac{4}{f_\pi}
\left\{
{\rm tr}((\del_\mu {a_1}_\nu-\del_\nu {a_1}_\mu)[\del^\mu\pi,\rho^\nu]
)\right. 
\\
\left.
+{\rm tr}((\del_\mu\rho_\nu-\del_\nu\rho_\mu)[\del^\mu\pi,a_1^\nu])
\right\}\ ,
\label{eq:L_a1pirho}
\end{split}
\end{gather}
where 
$
\rho^\mu\equiv \vec{\rho^\mu}\cdot\frac{\vec{\tau}}{2},\ 
\pi\equiv \vec{\pi}\cdot\frac{\vec{\tau}}{2}
$, and 
$
a_1^\mu\equiv \vec{a}_1^\mu\cdot\frac{\vec{\tau}}{2}\ .
$
The first Lagrangian gives the four-point Weinberg-Tomozawa (WT)
interaction and the second gives the three-point vertex of
$a_1\pi\rho$.

The advantage that we use the concept of the holographic QCD approach is that 
the large-$N_c$ condition ensures that the $a_1$ field in the Lagrangian
does not contain hadronic composite components, and hence we can avoid the
double-counting in the analysis of the mixing nature.  
The model contains two inputs, $f_\pi=92.4$ MeV and $m_\rho=776$ MeV,
giving the mass of the elementary $a_1$ meson $m_{a_1}=1189$ MeV and the
$a_1\pi\rho$ coupling constant $g_{a_1\pi\rho}=0.26$.
In the present study we consider the $a_1$ field 
appearing in the Lagrangian with these physical constants
to be the quark composite that
survives in the large $N_c$ limit. 
For the mass of the pion we
employ the physical  
value $m_\pi=138$ MeV that is isospin-averaged.
These interactions in Eqs.~(\ref{eq:L_WT}) and (\ref{eq:L_a1pirho}) are
essentially the same as those of the hidden-local
symmetry~\cite{Bando:1987br} except for the actual values of $m_{a_1}$ 
and $g_{a_1\pi\rho}$.  

%\subsection{kernels of Bethe-Salpeter equation}

Using these interactions, we take the WT potential $V_{\rm
WT}$~\cite{Roca:2005nm} and the $a_1$ pole term $V_{a_1}$ as,
\begin{gather}
V_{\rm WT}=\frac{\epsilon\cdot\epsilon'}{4f_\pi^2}
\{
3s -2(m_\rho^2+m_\pi^2)-\frac{1}{s}(m_\rho^2-m_\pi^2)^2
\},
\label{eq:V_WT}\\
V_{a_1}=-\frac{8}{f_\pi^2}g_{a_1\pi\rho}^2
\frac{\epsilon\cdot\epsilon'}{s-m_{a_1}^2+i\epsilon}(s-m_\rho^2)^2 \ ,
\label{eq:V_a1pole}
\end{gather}
after the $s$-wave projection with on-shell energies for external $\pi$
and $\rho$~\cite{Roca:2005nm}.
In actual calculations,
we should treat the polarization vectors
$\epsilon$ and $\epsilon'$ for 
the $\rho$ meson appropriately.  As reported in detail in
Ref.~\cite{Roca:2005nm}, by substituting the coefficient $v_{\rm
WT}(v_{a_1})$ defined by $V_{\rm WT}(V_{a_1}) \equiv -
\epsilon\cdot\epsilon' v_{\rm WT}(v_{a_1})$
for the potentials in Eq.~(\ref{eq:T-matrix}), we can obtain the
scattering amplitude for the transverse polarization mode, in which we
can find poles dynamically generated.

{In solving the scattering equation, we need to calculate the
propagator function $G$.}
  Because the potentials in Eqs.~(\ref{eq:V_WT}) and (\ref{eq:V_a1pole}) are
separable, the formal solution of Eq.~(\ref{eq:T_pirho}) becomes
algebraic, and the function $G$ 
%{contains a factor},
{is then given by,}
\begin{equation}
{G(\sqrt{s})=i}{\int\frac{d^4q}{(2\pi)^4}\frac{1}{(P-q)^2-m_\pi^2+i\epsilon}
\frac{1}{q^2-m_\rho^2+i\epsilon}\ ,}
\label{eq:G}
\end{equation}
{where $P$ is the total four-momentum as $P^2=s$.}
The {integral (\ref{eq:G})} diverges, and hence,
we regularize it by the 
dimensional regularization, and for the finite part we introduce
a subtraction constant $a(\mu)$ {as,
\begin{eqnarray}
G(\sqrt{s})&=&\frac{1}{16\pi^2}\left\{a(\mu)+\ln\frac{m_\rho^2}{\mu^2}
+\frac{m_\pi^2-m_\rho^2+s}{2s}\ln\frac{m_\pi^2}{m_\rho^2} \right.\nonumber\\
&+&\frac{q'}{\sqrt{s}}\left[\ln(s-(m_\rho^2-m_\pi^2)+2q'\sqrt{s})\right.\nonumber\\
&+&\ln(s+(m_\rho^2-m_\pi^2)+2q'\sqrt{s})\nonumber\\
&-&\ln(s-(m_\rho^2-m_\pi^2)-2q'\sqrt{s})\nonumber\\
&-&\left.\left.\ln(s+(m_\rho^2-m_\pi^2)-2q'\sqrt{s})-2\pi i
\right]
\right\},
\end{eqnarray}
where $\mu$ is the scale parameter that is set to be 900
MeV in this paper and $q'=\lambda^{1/2}(s,m_\rho^2,m_\pi^2)/2\sqrt{s}$.
}
A crucial point here is that the constant $a(\mu)$ is
chosen to be a {\it natural} value~\cite{Hyodo:2008xr} which ensures
that the resulting resonant states, if exist, are interpreted as a purely
hadronic composite.  
{Following the prescription in Ref.~\cite{Hyodo:2008xr}, we
choose $a(\mu)=-0.2$ at the renormalization scale $\mu$ which satisfies
the matching condition Re$\ G(\sqrt{s}=m_\rho)=0$. 
}
A choice of the subtraction constant $a{(\mu)}$
different from the natural value is equivalent to the 
introduction
of the
CDD (or elementary) pole that is not included in the model space of the scattering
problem~\cite{Hyodo:2008xr}. 

{
The polarization vector of the intermediate $\rho$ meson is treated in
the same way as reported in Ref.~\cite{Roca:2005nm}.
}

\section{Numerical results}
\subsection{nature of the $a_1$(1260) meson}
\begin{figure}[b]
\includegraphics[width=0.7\linewidth]{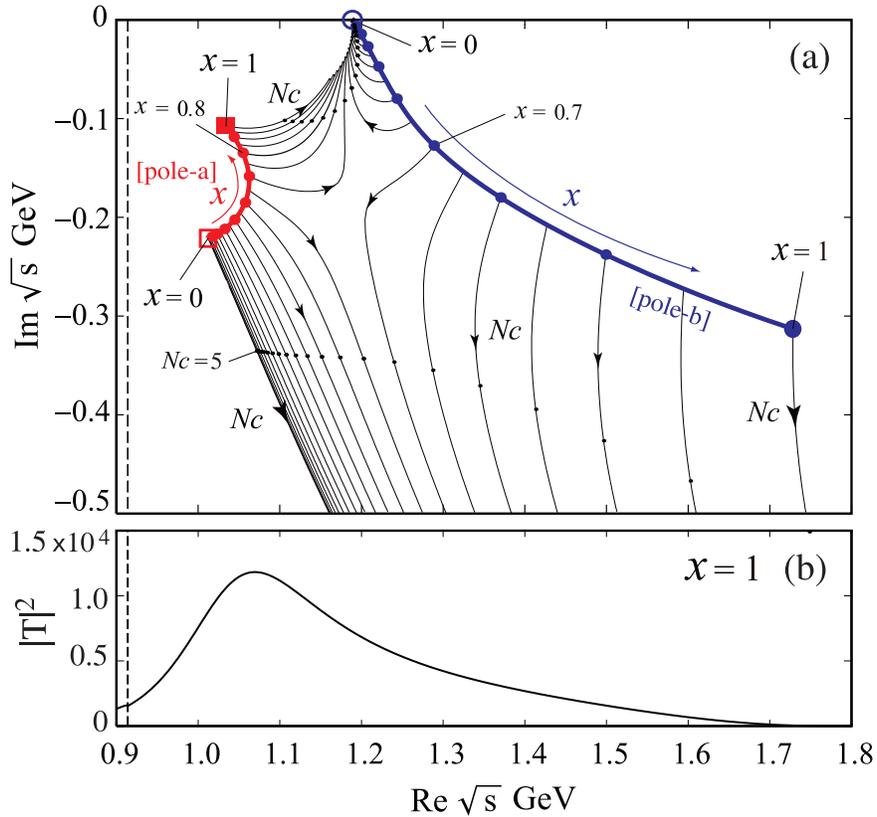}% Here is how to import EPS art
\caption{(Color online) (a) Trajectories of the poles in the full scattering amplitude
in Eq.~(\ref{eq:T-matrix}) by changing the mixing parameter $x$ (thick lines). 
The open square
 indicates the pole position of the composite $a_1$
and open circle
 indicates the elementary $a_1$ pole $(x=0)$.  The other end points of
 solid circle and square correspond to the physical points $(x=1)$.
Thin lines represent the pole-flows as $N_c$ is increased from
 $N_c=3$ for fixed $x$ with small dots at $N_c=5$.
The vertical dashed line denotes the $\pi\rho$
 threshold energy.  {(b) Squared amplitude {$|T|^2$} of $\pi\rho\rightarrow\pi\rho$
 process on the real energy axis {for} the mixing parameter $x=1$.}
\label{fig:pole_flow}
}
\end{figure}

Now we can evaluate the scattering amplitude  (\ref{eq:T-matrix}) or
(\ref{eq:T-matrix2}) numerically.  We find two poles at
(a) $\sqrt{s}=1033 - 107i$ MeV and at (b) $1728 -313i$ MeV,
corresponding to the physical states in the present model.  These pole
positions are significantly different from those of the two basis
states, $\sqrt{s_p}=1012-221i$ MeV and $m_{a_1}=1189$ MeV, because of
the mixing effect.  For further investigation, we vary the coupling
strength of $a_1\pi\rho$ by introducing a parameter $x (0 \le x \le 1)$,
$g_{a_1\pi\rho}\rightarrow xg_{a_1\pi\rho}$ which controls the mixing
strength.}

In Fig.~\ref{fig:pole_flow}(a), we
show the {resulting} pole-flow in the complex-energy plane  by changing the
mixing parameter $x$.
{At $x=0$, the poles corresponding to the basis states of the
composite and elementary $a_1$'s 
are found at the positions indicated by open square and open circle in
Fig.~\ref{fig:pole_flow}(a), respectively.} 
When the mixing is turned on, the pole starting from the
composite $a_1$ (we refer to it as ``pole-a'') 
approaches the real axis, ending at $1033-107i$ MeV when $x=1$ (solid square),
while that from the elementary $a_1$ pole (``pole-b'') goes far from the real axis 
and reaches
$1728-313i $ MeV when $x=1$ (solid circle).
%As we can see from the figure, 

In Fig.~\ref{fig:pole_flow}(b), we show the squared amplitude $|T|^2$
in Eq.~(\ref{eq:T-matrix}) (or (\ref{eq:T-matrix2})) at
$x=1$.  As shown in the figure, a peak structure is dominated 
by the pole-a, while a signal of the pole-b cannot be seen clearly
because of its huge width ($\sim 600$ MeV).
Therefore, the pole expected to be observed in experiments is the pole-a
located at lower energy position (solid square in
Fig.~\ref{fig:pole_flow}(a)) that comes from the  
composite $a_1$ pole.
Indeed, the contribution of the pole-b is found 
to interfere destructively with the tail of the pole-a
(around Re$\sqrt{s}\sim 1.7$--$1.8$ MeV) as shown in Fig.~\ref{fig:pole_flow}(b).
Our present model setting, however, is rather simple
and may not be applicable
{to} such a higher energy region where contributions from higher
coupled-channels neglected here {should be important.
A detailed investigation for an evidence of the second state of the
$a_1$ meson, namely the pole-b, is an interesting future work.

\begin{figure}[hbt]
\includegraphics[width=0.7\linewidth]{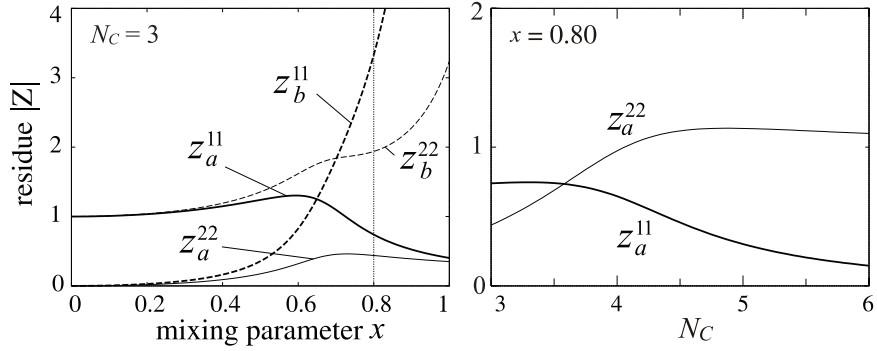}
\caption{Absolute value of the residues defined in
 Eq.~(\ref{eq:Res}). The left panel shows the mixing parameter $x$
 dependence at $N_c=3$ while the right panel is the $N_c$ dependence at
 {$x=0.8$}. The meaning of each line is indicated in the figure.
\label{fig:Res}}
\end{figure}

To study the mixing properties more quantitatively, we show the absolute values of the
residues as functions of the mixing parameter $x$ in Fig.~\ref{fig:Res} (left panel).
We can verify that,
before turning on the mixing, the {pole-a}
is purely the composite 
$a_1$ ($z_a^{11}=1$, $z_a^{22}=0$ at $x=0$) and
the {pole-b} is 
the elementary $a_1$ as we intended.
{One of the most important messages can be read from the magnitude
of $z_a^{11}$ and $z_a^{22}$ at $x=1$, which are the residues of
the possibly observed $a_1$ state.
While we should carefully discuss the meaning of the
residues of complex-value for resonant states, 
we can say that
the pole-a at $x=1$,
although its location is close to the composite $a_1$ pole, 
has a component of the elementary $a_1$ meson
{\em comparable to} that of the composite $a_1$.   This conclusion can be
drawn only after we look into the wave function by the residues $z$.} 

As for the pole-b, we find that the magnitude of the residues
$z_b^{22}$ and $z_b^{11}$ interchange at around $x\sim 0.7$, meaning
that the nature of the resonant state of {pole-b} changes from the
elementary particle-like structure to the composite one.  
For larger $x$ ($\gtrsim 0.8$), $z_b^{11}$ and $z_b^{22}$ become
larger than unity
because the energy dependence of the potentials become stronger at
the higher energy region where the pole-b is located.  {Once again, i}n
such a higher energy region (Re$\sqrt{s} \gtrsim 1.5$ GeV) our present
model may not be applicable.

\subsection{$N_c$ dependence and discussion on the classification test}
{Next, we test the large $N_c$ dependence of the pole positions
according to the scaling law of the pion decay constant $f_\pi$ as
$f_\pi\rightarrow f_\pi\sqrt{N_c/3}$.  In Fig.~\ref{fig:pole_flow}(a), we
also show the trajectories of the pole positions by changing the $N_c$ value
for fixed mixing strength.  At $x=0$, 
as reported in 
Ref.~\cite{Geng:2008ag}, we see the composite $a_1$ pole tends to have
a heavier-mass and a wider-width as $N_c$ is increased.
For small mixing parameter $x \lesssim 0.6$,
we 
find a similar behavior for the pole-a while the pole-b goes back to the elementary
$a_1$ position.
At $x\sim 0.7$, we find that the two $N_c$-trajectories {flip simultaneously},
and for $x \gtrsim 0.7$, the pole-b goes away from the real axis as the
composite $a_1$ does, while the pole-a approaches the elementary pole
position as $N_c$ is increased.
In this way, the result of the large $N_c$ classification depends 
strongly on the mixing parameter $x$, although the component 
of the composite $a_1$ is always larger than that of elementary 
($z_a^{11} > z_a^{22}$) at $N_c=3$ {as shown in
Fig.~\ref{fig:Res} (left panel)}. 

In Fig.~\ref{fig:Res} (right panel), we show the $N_c$ dependence of the 
residues of pole-a at $x = 0.8$. There we find that the magnitudes 
of the residues of the pole-a, $z_a^{11}$ and $z_a^{22}$, interchange at 
$N_c \sim 3.5$. This indicates that the nature of the resonance 
{\em changes as $N_c$ is varied}. Thus, for the mixed system of 
elementary and composite components, the large $N_c$ limit does
not always reflect the world at $N_c=3$. 
This is a consequence that there are two sources 
of the $N_c$ dependence, one from the $a_1\pi\rho$ three-point 
vertex which controls the mixing strength between the
two basis states, and the other from the WT interaction
which determines the pole position ($s_p$) of the basis state  
for the composite $a_1$. Their competition determines 
the nature of the $a_1$ states.
{Actually,
the $a_1 \pi\rho$ interaction Eq.~(\ref{eq:L_a1pirho}) is of order
$N_c^{-1/2}$ and vanishes in the large $N_c$ limit, where the two
states of $q \bar q$ meson and hadronic composite states decouple.   
In the real world of $N_c = 3$, however, the interaction remains finite
and gives an important contribution to the mixing dynamics as we showed.}
Therefore we conclude that and 
the large $N_c$ classification, which is
often used to identify the character of resonances, does not necessary
work for such a mixed systems.
%if it had an elemenary-like character as $N_c$ is increased.
%
%In Fig.~\ref{fig:Res} (right panel), we show the $N_c$ dependence of
%the residues of the pole-a at $x=0.8$.
%We find that the residue of the
%pole-a, $z_a^{11}$ and $z_a^{22}$ interchange
%at $N_c\sim 3.5$, 
%which indicates that the nature of resonance that we want to know 
%{\em changes as $N_c$ is varied.}
%Indeed, as for the pole-a at $x=0.8$,  although the component of the
%composite $a_1$ is larger than that of elementary ($z_a^{11}>z_a^{22}$)
%at $N_c=3$, in the complex-energy plane the pole-a approaches the
%elementary $a_1$ position as $N_c$ is increased as if it had an
%elementary-like structure.
%%
%This is a consequence of the $N_c$ dependence of the $a_1\pi\rho$ 
%three-point vertex that controls the mixing strength between two basis
%states, and also the $N_c$ dependence of the WT interaction that
%determines the pole position ($s_p$) of the basis state for the
%composite $a_1$. %%
%Therefore we conclude
%that the large $N_c$ procedure, which is often used to classify the
%character of resonances, does not necessarily work for the classification
%of the {mixed system} of a composite state and an elementary particle.
%That is to say,
%in such a mixed system, the large $N_c$ limit {\em does not always reflect the 
%world at $N_c=3$.}   
We will give general discussions of large-$N_c$ behavior
with two-level effective model separately in Ref.~\cite{Nawa}.

{
\section{Conclusion}
We {have} developed a general method to analyze the mixing structure of
hadrons consisting of two components of quark and hadronic composites.
{As an example,} the nature of
the $a_1(1260)$ axial-vector meson has been explored.
The present analysis points out theoretically that the $a_1$ meson 
has comparable amount of the elementary $a_1$ component to the
$\pi\rho$ composite $a_1$. 
Quest for evidences of the
mixing nature in physical observables is an interesting
future work.}

{
The method proposed in the present paper makes it possible to discuss
the validity of the 
classification {in} the large $N_c$ limit,
which is considered to
give a guiding principle to identify the nature of hadronic resonances.  
We have shown explicitly that the mixing nature of hadrons in the
large $N_c$ limit could differ from that at finite $N_c=3$.  We
conclude that the {simple} classification does not always work 
when admixture of components having different origins is
important. 
}

%{
%The
%discussions provided here  
%can be applied to various {hadronic} systems
%{especially to exotic systems}
%which are expected to be made up
%of multiple components having different origins. 
%}
%

{
\section*{Acknowledgement}
We express our thanks to T.~Hyodo for useful discussions. 
This work is in part supported by Grant-in-Aid for Scientific Research on
Priority Areas ``Elucidation of New Hadrons with a Variety of Flavors
(Nos.~22105510 (H.~N.), 22105509 (K.~N.), 22105507 (D.~J.) and
E01:21105006 (A.~H.))'' from the ministry of Education,  
Culture, Sports, Science and Technology of Japan.
This work was done in part
under the Yukawa International Program for Quark-hadron Sciences (YIPQS). 
}

\bibliography{a1_nature.bib}

\end{document}